\documentstyle[12pt]{article}
\newtheorem{theorem}{Theorem}
\begin{document}
\begin{titlepage}
\hspace{9cm} ULB--PMIF--93/01

\vspace{2.5cm}
\begin{centering}

{\huge Consistent couplings between fields with a gauge
freedom and deformations of the master equation}\\
\vspace{1cm}
{\large Glenn Barnich$^\dagger$ and Marc Henneaux$^*$\\
Facult\'e des Sciences, Universit\'e Libre de Bruxelles,\\
Campus Plaine C.P. 231, B-1050 Bruxelles, Belgium}\\
\end{centering}
\vspace{.25cm}
\begin{abstract}
The antibracket in BRST theory is known to define a map

\noindent $\rm{H^p \times H^q
\longrightarrow H^{p+q+1}}$ associating with two equivalence classes of BRST
invariant observables of respective ghost number p and q an equivalence class
of BRST invariant observables of ghost number p+q+1. It is shown that this map
is trivial in the space of all functionals, i.e., that its image contains only
the zeroth class. However it is generically non trivial
in the space of local functionals.

Implications of this result for the problem of consistent interactions among
fields with a gauge freedom are then drawn. It is shown that the obstructions
to constructing such interactions lie precisely in the image of the
antibracket
map and are accordingly inexistent if one does not insist on locality.
However consistent local interactions are severely constrained.
The example of the Chern-Simons theory is considered. It is proved that
the only consistent, local, Lorentz covariant interactions for the abelian
models are exhausted by the non-abelian Chern-Simons extensions.
\end{abstract}
\vspace{.25cm}
{\footnotesize ($^\dagger$)Aspirant au Fonds National de la Recherche
Scientifique (Belgium)\\
($^*$)Also at Centro de Estudios
Cient\'\i ficos de Santiago, Chile.}

\end{titlepage}
\def\carre{\vbox{\hrule\hbox{\vrule\kern 3pt
\vbox{\kern 3pt\kern 3pt}\kern 3pt\vrule}\hrule}}

\pagebreak

\section{Introduction}
The antifield formalism \cite{Batalin1,Batalin2} appears to be one of the most
powerful and elegant methods for quantizing arbitrary gauge theories.
Originally presented as a set of efficient working rules, its physical
foundations have been gradually clarified by showing how
gauge invariance is completely captured by BRST cohomology
\cite{Fisch1,Fisch2,Henneaux1}. Some of its geometrical aspects
(Schouten bracket, role of Stokes theorem in the proof of the gauge
independence
of the path integral) have been developed in \cite{Witten1} and more
recently in \cite{Henneaux2,Schwarz1,Schwarz2,Khudaverdian}. The somewhat
magic importance of
the antifield formalism in string field theory
\cite{Bochicchio,Thorn,Zwiebach,Witten2,Verlinde,Hata}
and its remarkable underlying algebraic structure
\cite{Stasheff,Getzler1,Lian,Getzler2,Penkava} have attracted further
considerable
attention (see also \cite{Alfaro}).
It is fair to believe that more interesting results are still to come.

The purpose of this letter is to reanalyze the long-standing problem of
constructing
consistent interactions among fields with a gauge freedom in the light of
the antibracket formalism.
We point out that this problem can be economically reformulated as a
deformation problem
in the sense of deformation theory \cite{Gerstenhaber}, namely that of
{\em{deforming
consistently the master equation}}. We then show, by using the properties of
the antibracket,
that there is no obstruction to constructing interactions that consistently
preserve
the gauge symmetries of the free theory if one allows the interactions to be
non local\footnote{These results are in line with the light-front analysis
of \cite{Bengtsson},
as well as with the work of \cite{Knecht} where the role of the master
equation is also
strongly stressed.}.
Obstructions arise only if one insists on locality. We provide a
reformulation of the deformation of
the master equation that takes locality into account, and illustrate the new
features
to which this leads by considering the three dimensional
Chern-Simons theory. We show that the only local, Lorentz covariant,
consistent
interactions for free
(abelian) Chern-Simons models are given by the non-abelian Chern-Simons
theories. We also
establish the rigidity of the non-abelian Chern-Simons models with a simple
gauge group.
A fuller account of our results will be reported elsewhere \cite{Barnich}.
\section{The master equation and the antibracket map}
We first recall some basic properties of the antifield formalism.
The starting point is the action in Lagrangian form $S_0 [\varphi^i]$, with
gauge
symmetries\footnote{We shall follow the presentation of the antifield
formalism
given in \cite{Henneaux2} (chapters 15, 17 and 18). We refer the reader to
that reference for more
information.}
\begin{equation}
\delta_\varepsilon \varphi^i = R^i_\alpha \varepsilon^\alpha .
\end{equation}
Given $S_0 [\varphi^i]$, one can, by introducing ghosts and antifields,
construct the solution
$S[\varphi^A,\varphi^*_A]$ of the master equation,
\begin{equation}
S = S_0 + \varphi^*_i R^i_\alpha C^\alpha + ...
\end{equation}
\begin{equation}
(S,S)=0
\end{equation}
where $\varphi^A\equiv (\varphi^i,C^\alpha,...)$ denotes collectively the
original fields, the
ghosts and the ghosts of ghosts if necessary, while $\varphi^*_A$ stands for
the antifields.
The solution $S$ of the master equation captures all the information about
the gauge structure
of the theory. The existence of $S$ reflects the consistency of the gauge
transformations.
The Noether identities, the (on-shell) closure of the gauge transformations
and the higher
order gauge identities are contained in the master equation $(S,S)=0$.
The original gauge invariant action $S_0$ itself and the gauge
transformations can be recovered
from $S$ by setting the antifields equal to zero in $S$ or in
${\delta S}/{\delta\varphi^*_i}$,
\begin{equation}
S_0=S[\varphi^A,\varphi^*_A =0]
\end{equation}
\begin{equation}
\delta_\varepsilon \varphi^i = {\delta S\over\delta\varphi^*_i}
[\varphi^A,\varphi^*_A =0]\ (C^\alpha
\longrightarrow\varepsilon^\alpha) .
\end{equation}

The BRST differential $s$ in the algebra of the fields and the antifields is
generated by
$S$ through the antibracket,
\begin{equation}
sA \equiv (A,S) .
\end{equation}
The BRST cohomology is denoted by $H^*(s)$. It is easy to verify that the
antibracket induces a
well defined map in cohomology,
\begin{equation}
(\cdot ,\cdot) : H^p(s)\times H^q(s)\longrightarrow H^{p+q+1}(s)\label{map}
\end{equation}
\begin{equation}
([A],[B]) = [(A,B)] \label{antibracketmap}
\end{equation}
where $[A]$ denotes the cohomological class of the BRST-closed element $A$.
We call (\ref{map}) ``the antibracket map". If one takes $[A] = [B]$
in (\ref{antibracketmap}), one gets
a map from $H^p(s)$ to $H^{2p+1}(s)$ sending $[A]$ on $[(A,A)]$.

It is sometimes useful to introduce auxiliary fields in a given theory,
namely, fields that can
be eliminated by means of their own equations of motion. This may, for
instance, simplify
the gauge structure and the geometric interpretation of the theory.
One then has various equivalent formulations and a natural question to ask
is :
what is the relationship between the BRST cohomologies and the antibracket
map of these
equivalent formulations~?
Not surprisingly, one has
\begin{theorem}
the BRST cohomologies $H^*(s)$ and $H^*(s^\prime)$ associated with two
formulations of a
theory differing in the auxiliary field content are isomorphic. Furthermore,
the isomorphism
i : $H^*(s)\longrightarrow H^*(s^\prime)$ commutes with the antibracket map.
\end{theorem}
{\em{Proof}} : the proof is direct and based on the explicit relationship
between the solutions
of the master equation of both formulations worked out in \cite{Henneaux3}.
We leave it as an exercise to the reader.

Using theorem 1, one can now establish the crucial result that the
antibracket map is
trivial\footnote{The proof assumes spacetime to be of the product form
$R\times M^{n-1}$ where $M^{n-1}$ is some $(n-1)$-dimensional spatial
manifold. It is also assumed
that the Lagrangian fulfills the standard regularity conditions that
guarantee
the existence of the reduced phase space, in terms of which the Cauchy
problem admits a unique
solution (see e.g. \cite{Henneaux2}). This implies in particular the
existence of proper
gauge fixings.}.
\begin{theorem}
the antibracket map is trivial, i.e., the antibracket of two BRST-closed
functionals
is BRST-exact.
\end{theorem}
{\em{Proof}} : the proof consists in two steps : {\em (i)} One adds auxiliary
fields
and fixes the gauge in such a way that {\em (a)} the gauge fixed equations of
motion
are of first order in the time derivatives and can be solved for
$\dot\varphi^A$ ;
and {\em (b)}~the BRST variation
of the fields depends only on the fields and not on their time derivatives
or on the antifields.
This can be done for instance
by going to the Hamiltonian formalism, and, as we have seen, modifies neither
the BRST
cohomology nor the antibracket map. {\em (ii)} By expressing the fields in
terms of initial
data on a Cauchy hypersurface,
one proves the existence, in each BRST cohomological class, of a
representative that does not
involve the antifields. More precisely, let $A[\varphi^A,\varphi^*_A]$ be a
solution of
$sA=0$ and let $\tilde A[\varphi^A]$ be the functional of the free initial
data that
coincides with $A[\varphi^A,\varphi^*_A=0]$ on-shell. One easily verifies
that $s\tilde A=0$.
Furthermore, $A$ and $\tilde A$ are in the same cohomological class due to
general properties
of the antifield formalism~\cite{Henneaux2}.
For representatives that do not involve the antifields, $(A,B)$ vanishes
identically and not just in cohomology. This proves the theorem
(A more detailed analysis will be
given
in \cite{Barnich}).
\section{Higher order maps}
The triviality of the antibracket map enables one to define higher order
operations in
cohomology. For example, if $[A] \epsilon H^p(s)$, one can define a squared
map
$H^p(s)\longrightarrow H^{3p+1}(s)$ as follows : the antibracket $(A,A)$ is a
coboundary.
Accordingly, there exist a functional $B$ of degree $2p$ such that
$(A,A) = (B,S)$.
The functional $B(\varphi,\varphi^*)$ is defined up to a cocycle.
Now $(A,B)$ is easily verified to be
BRST-closed and the cohomological class of $(A,B)$ does not depend on the
ambiguity in $B$.
Furthermore, $[(A,B)] = [(A^\prime,B^\prime)]$ if $[A] = [A^\prime]$.
Hence, the application $H^p(s)\longrightarrow H^{3p+1}(s)$ that maps $[A]$ on
$[(A,B)]$ is well-defined. In our case, however, the squared map and all the
other higher
order maps that can be defined in a similar fashion are trivial since one can
choose representatives in $H^p(s)$ such that
$(A,A)$ and hence $B$ both strictly vanish.
\section{Constructing consistent couplings as a deformation problem}
We now turn to the problem of introducing consistent interactions for a
``free" action
$\stackrel{(0)}{S_0} [\varphi^i]$ with ``free" gauge symmetries
\begin{equation}
\delta_\varepsilon \varphi^i = \stackrel{\smash{(0)}}{R}^i_\alpha
\varepsilon^\alpha ,
\end{equation}
\begin{equation}
{\delta\stackrel{(0)}{S}\over\delta\varphi^i}
\stackrel{\smash{(0)}}{R}^i_\alpha=0\ .
\end{equation}
We want to modify $\stackrel{(0)}{S_0}$
\begin{equation}
\stackrel{(0)}{S_0}\longrightarrow S_0 = \stackrel{(0)}{S_0} +
g \stackrel{(1)}{S_0} +
g^2 \stackrel{(2)}{S_0} + ...\label{fullaction}
\end{equation}
in such a way that one can consistently deform the original gauge symmetries,
\begin{equation}
\stackrel{\smash{(0)}}{R}^i_\alpha \longrightarrow R^i_\alpha =
\stackrel{\smash{(0)}}{R}^i_\alpha
+ g \stackrel{\smash{(1)}}{R}^i_\alpha +
g^2 \stackrel{\smash{(2)}}{R}^i_\alpha
+ ...\label{fullsymmetries} .
\end{equation}
By ``consistently", we mean that the deformed gauge transformations
$\delta_\varepsilon \varphi^i =
R^i_\alpha \varepsilon^\alpha$ are indeed gauge symmetries of the full
action (\ref{fullaction}),
\begin{equation}
{\delta(\stackrel{(0)}{S_0} + g \stackrel{(1)}{S_0} +
g^2 \stackrel{(2)}{S_0} + ...)\over\delta\varphi}
(\stackrel{\smash{(0)}}{R}^i_\alpha
+ g \stackrel{\smash{(1)}}{R}^i_\alpha + g^2 \stackrel{\smash{(2)}}
{R}^i_\alpha + ...)=0 \label{fullnoetheridentities} .
\end{equation}
This implies automatically that the modified gauge transformations close
on-shell
for the interacting action (see \cite{Henneaux2}, chapter 3).
In the case where the original gauge transformations are reducible,
one should also demand that
(\ref{fullsymmetries}) remain reducible. Interactions fulfilling these
requirements are called
``consistent". [It may be necessary to add further consistency requirements,
but this will not be considered here].

A trivial type of consistent interactions is obtained by making field
redefinitions
$\varphi^i\longrightarrow\bar{\varphi}^i = \varphi^i + g F^i + ...$  .
One gets
\begin{equation}
\stackrel{(0)}{S_0}\longrightarrow S_0 = \stackrel{(0)}{S_0}[\varphi^i +
g F^i + ...] =
\stackrel{(0)}{S_0} + g {\delta\stackrel{(0)}{S_0}\over\delta\varphi}F^i
+ ...\label{trivialinteractions}\ .
\end{equation}
Interactions that can be eliminated by field redefinitions are usually
thought of
as being no interactions. We shall say that a theory is rigid if the only
consistent deformations are
proportional to $\stackrel{(0)}{S_0}$ up to field redefinitions.
In that case, the interactions can be summed as
\begin{equation}
\stackrel{(0)}{S_0}\longrightarrow S_0 = (1 + k_1 g + k_2 g^2 + ...)
\stackrel{(0)}{S_0}
\end{equation}
and simply amount to a change of the coupling constant in front of the
unperturbed action.

The problem of constructing consistent interactions is a complicated one
because one must
simultaneously modify $\stackrel{(0)}{S_0}$ and $\stackrel{(0)}{R}^i_\alpha$
in such a way that
(\ref{fullnoetheridentities}) is valid order by order in $g$.
It has been studied for lower spins by many authors
(see for instance \cite{Arnowitt,Berends,Argyres} and references therein)
and some aspects of the
algebraic structure underlying the construction were clarified in \cite{Lada}.
One can reformulate more economically the problem in terms of the
solution $S$ of the master equation. Indeed, if the interactions can be
consistently
constructed, then the solution $\stackrel{(0)}{S}$ of the master equation
for the free theory
can be deformed into the solution $S$ of the master equation for the
interacting theory
\begin{equation}
\stackrel{(0)}{S}\longrightarrow S = \stackrel{(0)}{S} +
g \stackrel{(1)}{S} +
g^2 \stackrel{(2)}{S} + ...
\end{equation}
\begin{equation}
(\stackrel{(0)}{S},\stackrel{(0)}{S})=0\longrightarrow
(S,S)=0 \label{fullmasterequation}.
\end{equation}
The master equation $(S,S)=0$ guarantees that the consistency requirements
on $S_0$
and $R^i_\alpha$ are fulfilled.

There is a definite advantage in reformulating the problem of consistent
interactions as the
problem of deforming the master equation\footnote{Deforming the master
equation also appears
in renormalization theory where (\ref{fullmasterequation}) is replaced by
the equation
$(\Gamma,\Gamma)=0$ for the generating function of proper vertices
\cite{Zinn-Justin}.}.
It is that one can bring in the cohomological techniques of deformation
theory.
The master equation for $S$ splits according to the deformation parameter
$g$ as
\begin{eqnarray}
(\stackrel{(0)}{S},\stackrel{(0)}{S})&= 0 \label{deformation1}\\
2(\stackrel{(0)}{S},\stackrel{(1)}{S})&= 0 \label{deformation2}\\
2(\stackrel{(0)}{S},\stackrel{(2)}{S}) +
(\stackrel{(1)}{S},\stackrel{(1)}{S})&= 0 \label{deformation3}\\
&\vdots\ \ \ .\nonumber
\end{eqnarray}
The first equation is satisfied by assumption, while the second implies that
$\stackrel{(1)}{S}$
is a cocycle for the free differential $\stackrel{(0)}{s}\equiv
(\cdot,\stackrel{(0)}{S})$.
Suppose that $\stackrel{(1)}{S}$ is a coboundary, $\stackrel{(1)}{S}=
(\stackrel{(1)}{T},\stackrel{(0)}{S})$.
This corresponds to a trivial deformation because $\stackrel{(0)}{S_0}$
is then modified
as in (\ref{trivialinteractions})
\begin{equation}
\stackrel{(0)}{S_0}\longrightarrow \stackrel{(0)}{S_0} -
g {\delta\stackrel{(1)}{T}\over\delta\varphi^*_i}
{\delta\stackrel{(0)}{S}\over\delta\varphi^i}
\end{equation}
(the other modifications induced by $\stackrel{(1)}{T}$ affect the higher
order
structure functions which carry some intrinsic ambiguity \cite {Henneaux1}).
Hence, non trivial deformations are determined by the zeroth cohomological
space
$H^0(\stackrel{(0)}{s})$ of the undeformed theory. This space is generically
non-empty :
it is isomorphic to the space of observables \cite{Fisch1,Fisch2,Henneaux2}.

The next equation (\ref{deformation2}) implies that $\stackrel{(1)}{S}$
should be such that
$(\stackrel{(1)}{S},\stackrel{(1)}{S})$ is trivial in
$H^1(\stackrel{(0)}{s})$.
But we have seen that the map $H^0(\stackrel{(0)}{s})
\longrightarrow H^1(\stackrel{(0)}{s})$
induced by the antibracket is trivial and so, this requirement is
automatically satisfied.
Similarily, the higher order maps $H^0(\stackrel{(0)}{s})
\longrightarrow H^1(\stackrel{(0)}{s})$
are also trivial, which guarantees that the next terms $\stackrel{(3)}{S},
\stackrel{(4)}{S}, ...$
exist. Thus given an initial element $\stackrel{(1)}{S}$ of
$H^0(\stackrel{(0)}{s})$,
there is no obstruction in continuing the construction to get the complete
$S$.
The next terms $\stackrel{(2)}{S}, \stackrel{(3)}{S}, ...$ are determined up
to
an element of $H^0(\stackrel{(0)}{s})$, i.e., up to a gauge invariant
function.
At each order in $g$ there is the freedom of adding to the interaction an
arbitrary element
of $H^0(\stackrel{(0)}{s})$.

We can thus conclude that in the absence of particular requirements on the
form of the
interactions such as spacetime locality or manifest Lorentz covariance,
there is no obstruction to constructing interactions that preserve the
initial gauge symmetries
as in (\ref{fullnoetheridentities}).
In orther words, there is no ``no-go theorem".

\section{Spacetime locality of the deformation - The example of free abelian
Chern-Simons models}
The above construction does not yield, in general, a local action and is
somewhat formal.
In practice, it is usually demanded that the deformation be local in
spacetime, i.e.,
that $\stackrel{(1)}{S}, \stackrel{(2)}{S}, ...$ be {\em{local}}
functionals. This leads
to interesting developments.

In order to implement locality in the above analysis, we recall that if $A$
is a local functional
which vanishes for all allowed field configurations, $A=\int a =0$, then,
the $n$-form $a$
is a ``total derivative", $a=dj$,where $d$ is the spacetime exterior
derivative and $j$
is such that $\oint j =0$ (see e.g. \cite {Henneaux2} chapter 12). That is,
one can ``desintegrate"
equalities involving local functionals but the integrands are determined up
to $d$-exact terms.

Let $\stackrel{k}{S}=\int\stackrel{k}{\cal{L}}$ where
$\stackrel{k}{\cal{L}}$ is a n-form depending on the variables and a finite
number
of their derivatives, and let $\{a,b\}$ be the antibracket for such
$n$-forms, i.e.,
\begin{equation}
(A,B) = \int \{a,b\} \label{localantibracket}
\end{equation}
if $A = \int a$ and $B = \int b$. [Because $(A,B)$ is a local functional,
there exists $\{a,b\}$
such that (\ref{localantibracket}) holds, but $\{a,b\}$ is defined only up
to $d$-exact terms.
This ambiguity plays no role in the subsequent developments].
The equations (\ref{deformation1}-\ref{deformation3}) for $\stackrel{k}{S}$
read
\begin{eqnarray}
2\stackrel{(0)}{s}\stackrel{(1)}{\cal{L}}&=d\stackrel{(1)}{j}
\label{localdeformation}\\
\stackrel{(0)}{s}\stackrel{(2)}{\cal{L}} +
\{\stackrel{(1)}{\cal{L}},\stackrel{(1)}{\cal{L}}\}
=&d\stackrel{(2)}{j}\\
&\vdots\nonumber
\end{eqnarray}
in terms of the integrands $\stackrel{(k)}{\cal{L}}$.
The equation (\ref{localdeformation}) expresses that
$\stackrel{(1)}{\cal{L}}$ should
be BRST closed modulo $d$ and again, it is easy to see that a BRST-exact
term modulo $d$
corresponds to trivial deformations. Non trivial local deformations of the
master equation are
thus determined by
$H^0(\stackrel{(0)}{s}|d)$. [Note that an element of
$H^0(\stackrel{(0)}{s}|d)$ yields upon
integration
an element of $H^0(\stackrel{(0)}{s})$ only if appropriate surface terms
vanish. We shall not
investigate this question here and work with all the elements of
$H^0(\stackrel{(0)}{s}|d)$].

Now, while $(\stackrel{(1)}{S},\stackrel{(1)}{S})$ is always cohomologically
trivial,
it is not true, in general, that it is the BRST variation of a local
functional. Hence,
$\{\stackrel{(1)}{\cal{L}},\stackrel{(1)}{\cal{L}}\}$ {\em{may not}}
be BRST-exact modulo $d$, and the map
\begin{equation}
H^p(\stackrel{(0)}{s}|d) \times H^q(\stackrel{(0)}{s}|d) \longrightarrow
H^{p+q+1}(\stackrel{(0)}{s}|d)
\end{equation}
defined by the antibracket appears to possess a lot of structure.
Furthermore, even when the image of
$\{\stackrel{(1)}{\cal{L}},\stackrel{(1)}{\cal{L}}\}$
is trivial in $H^0(\stackrel{(0)}{s}|d)$, so that the squared map
$H^0(\stackrel{(0)}{s}|d)\longrightarrow H^1(\stackrel{(0)}{s}|d)$ can be
defined,
there is no guarantee that this squared map is trivial. For this reason,
the
construction of local, consistent interactions is a problem that is quite
constrained.

To illustrate this point, we shall analyze the case of the abelian
Chern-Simons models in three
dimensions.

The action is given by
\begin{equation}
\stackrel{(0)}{S_0}=\int d^3x {1\over 2}\varepsilon^{ijk}k_{ab}A^a_iF^b_{jk}
\end{equation}
where $k_{ab}$ is a non degenerate, symmetric and constant matrix.
The equations of motion imply $F^a_{ij}=0$.
An irreducible set of gauge transformations can be taken to be
\begin{equation}
\delta_{\varepsilon} A^a_i=\partial_i\varepsilon^a\ .
\end{equation}
The minimal solution to the classical master equation is
\begin{equation}
\stackrel{(0)}{S}=\stackrel{(0)}{S_0}+\int d^3x A^{i*}_a \partial_i C^a
\end{equation}
and the local version of the BRST symmetry is then
\begin{equation}
\stackrel{(0)}{s}=\varepsilon^{ijk}F^b_{jk}k_{ba}\stackrel{\rightarrow}{
{\partial\over\partial A^{i*}_a}}
-\partial_i A^{i*}_a \stackrel{\rightarrow}{{\partial\over\partial C^*_{a}}}
+\partial_i C^a
\stackrel{\rightarrow}{{\partial\over\partial A^{a}_i}}
\end{equation}
with $[\stackrel{(0)}{s},\partial_i]=0$ and $\stackrel{(0)}{s} d + d
\stackrel{(0)}{s}=0$.
As we have pointed out, the perturbation $\stackrel{(1)}{\cal{L}}$
should obey
(\ref{localdeformation}),
\begin{equation}
\stackrel{(0)}{s}\stackrel{(1)}{\cal{L}}+da_{[2]}=0\label{localtop}
\end{equation}
i.e., should define an element of $H^0(\stackrel{(0)}{s}|d)$. The
equation (\ref{localtop})
can be analyzed along lines familiar from the algebraic study of anomalies.
Indeed, one gets from (\ref{localtop}) a set of ``descent equations"
\cite{Stora,Zumino}
\begin{eqnarray}
\stackrel{(0)}{s}\stackrel{(1)}{\cal{L}}+da_{[2]}=0\label{topdescent}\\
\stackrel{(0)}{s} a_{[2]}+da_{[1]}=0\\
\stackrel{(0)}{s} a_{[1]}+da_{[0]}=0\\
\stackrel{(0)}{s} a_{[0]}=0\label{bottomdescent}\ .
\end{eqnarray}
To solve (\ref{localtop}), one needs to find the most general element at
the bottom
of the ladder that can be lifted all the way up to yield an element of
$H(\stackrel{(0)}{s}|d)$. This is the procedure followed in
\cite{Dubois-Violette2}.
Now, the last element of a descent belongs to $H(\stackrel{(0)}{s})$, and
must be a polynomial
in the ghosts $ C^a $ . [Because the equations of motion imply $F^a_{ij}=0$,
$F^a_{ij}$ is trivial in cohomology].
Thus
\begin{equation}
a_{[0]}=f_{abc}C^a C^b C^c
\end{equation}
where
$f_{abc}$ is completely antisymmetric. This implies
\begin{equation}
a_{[1]}=3f_{abc}A^a C^b C^c + m_{ab}C^a C^b
\end{equation}
where $m_{ab}C^a C^b$ belongs to $H(\stackrel{(0)}{s})$ and $m_{ab}$ is a
constant $1$-form.
By Lorentz covariance, this term must be zero.
This leads then to
\begin{equation}
a_{[2]}=-{3\over 2}f^a_{bc}\ ^* A^{*}_a C^b C^c+
3f_{abc}A^a\wedge A^b C^c
\end{equation}
and finally to
\begin{equation}
\stackrel{(1)}{\cal{L}}={1\over 6}f^a_{bc}\ ^* C^*_a C^b C^c +
f^a_{bc}3\ ^* A^{*}_a \wedge A^b C^c +f_{abc}A^a\wedge A^b\wedge A^c \ .
\end{equation}

It should be noted that $\stackrel{(1)}{S}=\int\stackrel{(1)}{\cal{L}}$ is
$\stackrel{(0)}{s}$-trivial in the space of all functionals.
Indeed, assuming that the fields decrease at infinity{\footnote
{Different boundary conditions or a non trivial spacetime topology
would require a more sophisticated treatement.}}, one can decompose
$A^a_i$ and $A^{i*}_a$
as
\begin{equation}
A^a_i=\partial_i\varphi^a + A^{Ta}_i,\ \ A^{i*}_a=\partial^i\varphi_a^* +
\varepsilon^{ijk}
\partial_j\mu^*_{ka}
\end{equation}
Because $A^{Ta}_i=0$ by the equations of motion, one finds that
$\stackrel{(1)}{S}_0$
vanishes on-shell,
\begin{equation}
\stackrel{(1)}{S}_0\approx \int {2\over 3}\partial_i\varphi^a
\partial_j\varphi^b
\partial_k\varphi^c
\varepsilon^{ijk}f_{abc}d^3x =0 \ .
\end{equation}
This implies that $\stackrel{(1)}{S}$ is BRST-exact (\cite{Henneaux2})
and indeed
\begin{equation}
\stackrel{(1)}{S}=(F,\stackrel{(0)}{S})\label{exactness}
\end{equation}
with
\begin{eqnarray}
F=\stackrel{(0)}{s}\int d^3xf^a_{bc}(({1\over 6}
\mu^*_{ia} A^{Tb}_j A^{Tc}_k +
{1\over 2} \mu^*_{ia} \partial_j\varphi^b \partial_k\varphi^c +\nonumber\\
\partial_i\mu^*_{ja} A^{Tb}_k \varphi^c)\varepsilon^{ijk}-
\carre^{-1}\partial^i
C^*_a A^{Tb}_i C^c -\carre^{-1}\partial^i
C^*_a \partial_j\varphi^b C^c)\ .
\end{eqnarray}
However, $F$ is a non-local functional of the fields and thus,
$\stackrel{(1)}{S}$
cannot be eliminated by local redefinitions. One then computes
$(\stackrel{(1)}{S},\stackrel{(1)}{S})$. One finds
\begin{eqnarray}
(\stackrel{(1)}{S},\stackrel{(1)}{S})=8\int d^3x
\{(f_{abc}\varepsilon^{ijk}A^b_j A^c_k
+f^b_{ac} A^{i*}_b C^c )(+f^a_{de} A^d_i C^e)+\nonumber\\
f^a_{bc}( A^{i*}_a A^b_i  + C^*_a C^b) ({1\over 2}f^c_{de} C^d C^e))\}\ .
\end{eqnarray}
The integrand of this expression is a $\stackrel{(0)}{s}$-cocycle modulo
$d$ because the Jacobi
identity for the local antibracket holds modulo $d$.
In order to construct a non trivial local interaction, this cocycle must
be trivial in
$H(\stackrel{(0)}{s}|d)$.
Because the image of $\stackrel{(0)}{s}$ and $d$ contains no terms without
derivatives,
a necessary and sufficient condition for this cocycle to be
$\stackrel{(0)}{s}$-trivial
modulo $d$ is that it vanishes. This is the case if and only if the
constants $f^a_{bc}$ verify the Jacobi identity, even though
$(\stackrel{(1)}{S},\stackrel{(1)}{S})$ is BRST-exact in the space of all
functionals for arbitrary choices of $f^a_{bc}$ thanks to
(\ref{exactness}).
This implies that $\stackrel{(0)}{S} + \stackrel{(1)}{S}$ is a solution to
our deformation problem
which corresponds of course to the well-known non-abelian
Chern-Simons theories~:
\begin{equation}
\stackrel{(0)}{S_0} + \stackrel{(1)}{S_0} =
\int d^3x ({1\over 2}\varepsilon^{ijk}k_{ab}A^a_iF^b_{jk} +
{2\over 3}\varepsilon^{ijk}f_{abc}A^a_i A^b_j A^c_k )
\end{equation}
Accordingly, the only consistent, Lorentz covariant couplings of
abelian Chern-Simons
models are the non-abelian extensions.
\section{Rigidity of non-abelian Chern-Simons theory}
We close this letter by proving the rigidity of the Chern-Simons theory
with a simple gauge group.
The descent equations for $\stackrel{(1)}{\cal{L}}$
are identical to (\ref{topdescent})-(\ref{bottomdescent}),
but this time, the $0$-form $a_{[0]}$ of ghost number $3$
should be closed for the non-abelian BRST differential.
The only non trivial element of $H^3(\stackrel{(0)}{s})$ is the primitive
form
$\alpha tr C^3$, where $\alpha$ is a priori an invariant polynomial in the
non-abelian
field strength $F$, which, however can be set equal to zero, since the
equations of motion are $F_{ij}=0$.
The primitive form $\alpha tr C^3$ can be lifted as in the familiar
Yang-Mills case \cite{Dubois-Violette2}
to yield $\alpha$ times the Chern-Simons action.
Since there is no cohomology in ghost degree $2$, $1$ or $0$ (apart from the
irrelevant constants),
there is no other element that can be lifted to yield another solution
from a shorter descent.
This proves the rigidity of the Chern-Simons action.

\section{Conclusion}
Reformulating the problem of consistent interactions in terms of
deformations of the
master equation allows the use of powerful BRST cohomological techniques.
The triviality of the antibracket map in cohomology in the space of all
functionals allows
to built consistent interactions from any gauge invariant functionals of the
undeformed theory. However, these interactions may be non local and
obstructions
on consistent local couplings do exist in practise.
The study of these obstructions require additional tools familiar
from the study of anomalies.
The analysis has been
illustrated for the Chern-Simons models.

\section{Acknowledgements}
We acknowledge fruitful discussions with Michel Dubois-Violette,
Marc Knecht, Jim Stasheff and Claudio Teitelboim. This work has been
supported in
part by the ``Fonds National de la Recherche Scientifique (Belgium)" and
by a research
contract with the Commission of the European Communities.
\pagebreak

\font\pet=cmr10 at 10truept
\font\bf=cmbx10 at 10truept

\pet

\end{document}